\documentstyle[preprint,pre,aps,eqsecnum]{revtex}

\begin{document}
\draft
\tightenlines
\preprint{
\rightline{\vbox{\hbox{\rightline{McGill/98-44}}}}}


\title{Splintering central nuclear collisions with a
momentum-dependent lattice Hamiltonian theory}

\author{Declan Persram and Charles Gale}

\address{ Physics Department, McGill University\\
         3600 University St., Montr\'eal, QC, H3A 2T8, Canada\\}

\date{\today}

\maketitle
\begin{abstract}
We formulate a lattice Hamiltonian model for intermediate
energy heavy ion collisions. Our approach incorporates a
momentum-dependent nuclear mean field that yields an optical
potential that agrees with proton-nucleus experiments and that also receives
support from measurements involving heavy ions. We
emphasize the precision of our numerical results in connection
with energy and momentum conservation. We first outline the static
properties of our solutions, after which we consider recent stopping
power results for heavy ion collisions. The results obtained with
different types of nuclear mean fields are compared with each other. 
Consequences of the stopping power data on the determination of the 
nuclear equation of state are outlined.
\end{abstract}



\section{Introduction }
Heavy ion collisions remain the only practical area of research which
addresses the physics of strongly interacting matter in states far away
from equilibrium. As such, it represents a rich and challenging field
well worthy of
intellectual pursuit and  is fundamental to a deeper  and a more
complete understanding of Nature. At high energies, a goal of this
program is to form and study a new state of matter which is a prediction
of QCD: the quark-gluon plasma~\cite{mey96}. At lower energies, the
experimental and theoretical efforts have focused on the need to
characterize and quantify the nuclear equation of state~\cite{nunu97}.
This physics also has an important
role to play in the theory of supernov{\ae} and that of neutron star
properties~\cite{pra97}.  It is that energy regime that we consider
in this work.

A popular practice consists of characterizing the nuclear equation
of state in terms of its coefficient of compressibility, K. This
number can be deduced from Hartree-Fock-plus-RPA analyses of giant
monopole resonances in finite nuclei~\cite{bla76}. In heavy ions, 
in order to
identify novel many-body features without ambiguity it is
imperative  to provide a realistic model of the nuclear
reaction dynamics. An approach that has proven to be extremely
successful is the Boltzmann-Uehling-Uhlenbeck (BUU) model of heavy
ion collisions~\cite{bdg88}. In BUU simulations, nucleons can
suffer two-body collisions and can also move on curved trajectories
owing to interactions with the self-consistent mean field. The
interaction we use in this work is described in Refs.~\cite{MDYI,zhang}.
A large body of work in intermediate energy
nuclear collisions has been devoted to the measurement and to the
theoretical calculation of nuclear flow~\cite{dan85,gut89,eos95,had95}.
In connection with this observable, it is fair to state that at
present little compelling evidence exists from intermediate energy
heavy ion data to suggest that the value of K derived from heavy
ion collisions \cite{zhang} should be different from that inferred 
from recent giant
monopole studies~\cite{monop}. Both methods of analysis suggested K
$\approx$ 215 MeV. However, a recent study can be found in Ref.
\cite{YCL99}. 

Near the low end of the intermediate energy spectrum, some studies
have put forward the possibility of observing experimental
signatures of new phenomena. A good example is that of reduced
in-medium nucleon-nucleon  cross 
sections~\cite{bauer,klak,hombach,mota92}. While the
confirmation of such manifestations would indeed be extremely
interesting, one must keep in mind that such ``new physics''
issues must be addressed with an approach that incorporates all of
the known physics in a computationally tractable model. Our goal
in this paper is to present such a model. At beam energies such as
the ones under consideration in this work, the problem of energy
and momentum conservation in transport models is a pressing one.
Also, it is of course necessary to include the Coulomb interaction for
adequate phenomenology. Finally, the momentum dependence of the
nuclear mean field is an unavoidable feature, both from the point
of view of theory \cite{jeu,mafl} as from that of
experiment~\cite{optmod0,optmod1}.
Those elements are incorporated in our model. The
conservation law requirements vital in this low energy environment 
are enforced via the Lattice
Hamiltonian algorithm~\cite{lenk_pand}.

Our paper is organized as follows: the next section introduces our
Lattice Hamiltonian. The following section explores its static
solutions. We then proceed to a discussion  of the Vlasov limit
of our transport theory, followed by a full BUU analysis. We 
apply our model to a specific measurement of stopping power in
nuclear collisions. We close with a summary and conclusion.

\section{Lattice Hamiltonian}
We wish to solve the equations of motion for a system of particles
interacting via a self-consistent mean field potential.  The first
step in our numerical solution is to write down the Hamiltonian
for a system of particles embedded in a six dimensional
lattice ($\delta x,\delta p$) in configuration and momentum space.
The phase space distribution function of a single particle (Roman index
``$i$'')
located at configuration (momentum) space lattice site
$\alpha\ (\pi)$ is
\begin{eqnarray}
    f_{\alpha\pi}^{i}=R(\vec{r}_{\alpha}-\vec{r}_{i})
              P(\vec{p}_{\pi}-\vec{p}_{i}).
\end{eqnarray}
In the above, $R$ ($P$) is the configuration (momentum) space form factor which
we will specify shortly.
If we simultaneously consider $N_{ens}$ such systems and wish our distribution
function to represent the average of these systems, then the distribution
function must be normalized such that
\begin{equation}
    (\delta x)^{3}(\delta p)^{3}\sum_{\alpha\pi}f_{\alpha\pi}^{i}
    =N_{ens}^{-1}.
\label{phase_space_norm}
\end{equation}
We define a ``test particle'' as one of the particles from the
$N_{ens}$ systems defined above. Note that summing the left hand side of
Eq. \ref{phase_space_norm} over ``$i$'', one obtains the total number
of nucleons: the proper
normalization. Now before we can write down the
total Hamiltonian, we need a form for the 
potential energy density. One choice used for heavy ion
calculations which takes into account non local mean field effects
is known as the MDYI potential~\cite{MDYI}. This interaction yields 
good bulk nuclear matter and optical potential
properties \cite{optmod1}. It also generates successful heavy ion phenomenology
\cite{zhang,gale90}.   This potential energy density reads:
\begin{eqnarray}
    V(\vec{r})=
    \frac{A}{2}\frac{\rho(\vec{r})^{2}}{\rho_{0}}
    +\frac{B}{\sigma+1}\frac{\rho(\vec{r})^{\sigma+1}}{\rho_{0}^{\sigma}}
    +\frac{C\Lambda^{2}}{\rho_{0}}\int\int d^{3}\!p\,d^{3}\!p'
    \frac{f(\vec{r},\vec{p})f(\vec{r},\vec{p}\,')}
         {\Lambda^{2}+(\vec{p}-\vec{p}\,')^{2}}\ ,
\end{eqnarray}
where $f$ is the phase space distribution function, and $\rho_0$ is the
density of nuclear matter. 
However since we are now working on a lattice, we require the discretized version
of the above:
\begin{eqnarray}
    V_\alpha=
    \frac{A}{2}\frac{\rho_{\alpha}^{2}}{\rho_{0}}
    +\frac{B}{\sigma+1}\frac{\rho_{\alpha}^{\sigma+1}}{\rho_{0}^{\sigma}}
    +\frac{C\Lambda^{2}}{\rho_{0}}(\delta p)^{6}
    \sum_{\pi\pi'}\frac{f_{\alpha\pi}f_{\alpha\pi'}}
               {\Lambda^{2}+(\vec{p}_{\pi}-\vec{p}_{\pi'})^{2}}.
\end{eqnarray}
The five constants $A,B,C,\sigma$ and $\Lambda$ in $V_{\alpha}$  were chosen to
reproduce the ground state binding energy per nucleon $E/A(\rho_{0})=-16$ MeV,
the nuclear compressibility K=215 MeV, the zero pressure condition
$P(\rho_{0},T=0)=0$, and the real part of the single particle potential
(see below) to satisfy
$U(\rho_{0},p=0)=-75$ MeV and
$U(\rho_{0},p^{2}/2m=300$ MeV$)=0$. Our parameters yield
$U(\rho_{0},p\rightarrow\infty)=30.5$ MeV and the effective mass at the Fermi
surface is $m^{*}/m=0.67$.
This parameter set as well as others investigated in this work are shown in
Table \ref{table1}.
Note that in the above $f_{\alpha\pi}\!\equiv\!\sum_{i}f_{\alpha\pi}^{i}$ and the
configuration space density at site $\alpha$ is  
$\rho_{\alpha}\!=\!(\delta p)^{3}\sum_{\pi}f_{\alpha\pi}$.  The total potential
energy density at site $\alpha$ shown in the above discretized version of 
$V_\alpha$ is due to all of the test particles.  To obtain
the single particle potential we must unfold the phase space
distribution function from
$V_{\alpha}$. We obtain the single particle potential from the above potential
energy density through the functional derivative \cite{negele} 
\begin{eqnarray}
    U_{\alpha\pi}&\equiv&\frac{\delta V_{\alpha}}{\delta f_{\alpha\pi}}
	\\
    &=&A\frac{\rho_{\alpha}}{\rho_{0}}
    +B\frac{\rho_{\alpha}^{\sigma}}{\rho_{0}^{\sigma}}
    +\frac{2C\Lambda^{2}}{\rho_{0}}(\delta p)^{3}\sum_{\pi'}
    \frac{f_{\alpha\pi'}}{\Lambda^{2}+(\vec{p}_{\pi}-\vec{p}_{\pi'})^{2}}.
\label{u_discrete}
\end{eqnarray}
Before we move any further, a simplification will be made to the lattice phase
space density. We will consider the momentum space lattice to have a
spacing of $\delta p\!=\!0$; a continuum in momentum space. With 
this, our phase space density is
\begin{eqnarray}
    f_{\alpha\pi}^{i}=R(\vec{r}_{\alpha}-\vec{r}_{i})
              \delta(\vec{p}_{\pi}-\vec{p}_{i}),
\end{eqnarray}
where $R$ may or may not be a delta function. With this form factor,
our potential energy density and single particle potential read:
\begin{equation}
    V_\alpha=
    \frac{A}{2}\frac{\rho_{\alpha}^{2}}{\rho_{0}}
    +\frac{B}{\sigma+1}\frac{\rho_{\alpha}^{\sigma+1}}{\rho_{0}^{\sigma}}
    +\frac{C\Lambda^{2}}{\rho_{0}}
    \sum_{ij}\frac{R(\vec{r}_{\alpha}-\vec{r}_{i})
               R(\vec{r}_{\alpha}-\vec{r}_{j})}
              {\Lambda^{2}+(\vec{p}_{i}-\vec{p}_{j})^{2}}
\label{discretized_v_mdyi}
\end{equation}
\begin{equation}
    U_\alpha(\vec{p})=A\frac{\rho_{\alpha}}{\rho_{0}}
        +B\frac{\rho_{\alpha}^{\sigma}}{\rho_{0}^{\sigma}}
        +\frac{2C\Lambda^{2}}{\rho_{0}}
    \sum_{i}\frac{R(\vec{r}_{\alpha}-\vec{r}_{i})}
             {\Lambda^{2}+(\vec{p}-\vec{p}_{i})^{2}}\ .
\label{discretized_u_mdyi}
\end{equation}
Now from~(\ref{phase_space_norm}) and~(\ref{discretized_v_mdyi})
we obtain the total Hamiltonian of all test particles ($N_{ens}$
systems of $A$ nucleons). With the Hamiltonian 
\begin{equation}
    H=\sum_{j}^{A\times N_{ens}}\frac{p_{j}^{2}}{2m}
    +N_{ens}(\delta x)^{3}\sum_{\alpha}V_{\alpha} \ , 
\label{total_hamiltonian}
\end{equation}
we can write down
the equations of motion for the $i^{th}$ test particle:
\begin{eqnarray}
    \dot{\vec{r}_{i}}&=&
            \nabla_{\vec{p}_{i}}H
            =\frac{\vec{p}_{i}}{m}
            +N_{ens}(\delta x)^{3}\sum_{\alpha}
             R(\vec{r}_{\alpha}-\vec{r}_{i})
             \nabla_{\vec{p}_{i}}U_{\alpha}(\vec{p}_{i})
\label{ham_x_latt}
	\\
    \dot{\vec{p}_{i}}&=&
            -\nabla_{\vec{r}_{i}}H
            =-N_{ens}(\delta x)^{3}\sum_{\alpha}
	     U_{\alpha}(\vec{p}_{i})
             \nabla_{\vec{r}_{i}}R(\vec{r}_{\alpha}-\vec{r}_{i})\ .
\label{ham_p_latt}
\end{eqnarray}

Up until now we have left the exact form of the configuration space form
factor unspecified. There is however a special case which we will consider. If
$R$ does not contain a delta function,
then~(\ref{ham_x_latt}) and (\ref{ham_p_latt}) give us the 
``Lattice Hamiltonian'' equations of motion. Suppose
however that we let $\delta x\!=\!0$ so that we have a continuum in both
configuration and momentum space. The normalized (see
equation~\ref{phase_space_norm}) phase space distribution function takes the
following form:
\begin{equation}
    f_{\alpha\pi}^{i}=N_{ens}^{-1}\delta(\vec{r}_{\alpha}-\vec{r}_{i})
              \delta(\vec{p}_{\pi}-\vec{p}_{i}).
\label{continuum_phase_space}
\end{equation}
Now, if we insert the above into~(\ref{ham_x_latt}) and 
(\ref{ham_p_latt}) we get a new set of equations of
motion which read
\begin{eqnarray}
        \dot{\vec{r}_{i}}&=&\frac{\vec{p}_{i}}{m}
             +\nabla_{\vec{p}_{i}}U(\vec{r}_{i},\vec{p}_{i})
\label{ham_x}
	\\
        \dot{\vec{p}_{i}}&=&-\nabla_{\vec{r}_{i}}U(\vec{r}_{i},\vec{p}_{i}).
\label{ham_p}
\end{eqnarray}
For relativistic kinematics, the $m$ above can be replaced by 
$\sqrt{p^{2}+m^{2}}$.
If the test particles evolve according to Hamilton's equations, their
phase space density, $f$, will satisfy the Vlasov equation. 
This 
equation can also be identified with the time evolution of 
the Wigner transform of
the one-body density matrix (which in turn is directly related to
the many body wave function of a system of particles) \cite{bdg88}. It is
\begin{eqnarray}
    \frac{\partial}{\partial t}f(\vec{r},\vec{p})
    +\nabla_{\vec{p}}H\cdot\nabla_{\vec{r}}f(\vec{r},\vec{p})
    -\nabla_{\vec{r}}H\cdot\nabla_{\vec{p}}f(\vec{r},\vec{p})
    =0.
\label{vlasov_equation}
\end{eqnarray}
From here on we refer
to equations~(\ref{ham_x_latt}) and (\ref{ham_p_latt}) as 
``LHV'' (Lattice Hamiltonian Vlasov) and 
equations~(\ref{ham_x}) and (\ref{ham_p}) 
as ``TPV'' (test particle Vlasov).

We have mentioned that~(\ref{ham_x_latt}) and (\ref{ham_p_latt}) 
are the equations of motion for a
phase space density without configuration space delta functions but we did not
offer a specific choice for the form factor. We follow the work of Lenk and
Pandharipande~\cite{lenk_pand} and adopt the form factor 
\begin{eqnarray}
    R(\vec{r}_{\alpha}-\vec{r}_{i})&=&\frac{1}{N_{ens}(n\delta x)^{6}}
                    g(x_{\alpha}-x_{i})
                    g(y_{\alpha}-y_{i})
                                        g(z_{\alpha}-z_{i})
	\\
    g(x_{\alpha}-x_{i})&\equiv&(n\delta x-|x_{\alpha}-x_{i}|)
                \Theta(n\delta x-|x_{\alpha}-x_{i}|)\ .
\end{eqnarray}
The normalization condition~(\ref{phase_space_norm}) is satisfied by the
above and ``$n\delta x$'' is the effective geometric radius of a test particle.
We choose two sets of parameters, $n\!=\!1,\delta x\!=\!1.50$ fm; and 
$n\!=\!2,\delta
x\!=\!0.75$~fm. These are listed in Table \ref{table2}.  Often, when 
one wishes to simulate interactions of heavy
ions the TPV method is employed, and one 
utilizes a finite configuration space grid. 
One then resolves the test particles up to some Euler grid scale and then 
assumes the validity of the TPV equations. This is clearly not correct and as
we shall see leads to violation of energy conservation which can be quite
extreme in some cases. Since this is a popular method, we too
will artificially smooth the configuration space and introduce a non zero
lattice constant when solving for TPV. 
One important aspect which differentiates the two methods is intimately
connected with the two sets of equations of motion used. In particular the LHV
equations of motion depend explicitly on the exact positions of all 
the particles within the cells. On the other hand, in the TPV method
the test particles are only resolved up to a cell constant. 

Note that the lattice Hamiltonian method with a MDYI-type 
momentum-dependent nuclear
potential has been previously used by us in the context of nuclear 
flow inversion \cite{dec}.

\section{Ground State Nuclei}
The first step in simulating a collision between two heavy ions is to
initialize the ground state or starting position of the two nuclei. 
We should thus be able to approximately reproduce the binding energy
per nucleon inside a nucleus. In introducing the nuclear matter
potential in the previous section we have neglected two ingredients
of the mean field that are important for practical applications. The 
first is the long range Coulomb potential and the 
second is the symmetry energy which we call here the 
isospin potential. With these, the potential energy density should read:
\begin{eqnarray}
    V_{\alpha}=V_{\alpha}^{nuc}+V_{\alpha}^{coul}+V_{\alpha}^{iso}
\end{eqnarray}
Here, $V_{\alpha}^{nuc}$ can be the MDYI potential from the previous section or any
other potential one wishes to consider.
The Coulomb potential takes its usual form and for the isospin, we adopt a form
for the single particle potential previously used~\cite{mbtsang,b_li_1} which
has the following potential energy density:
\begin{eqnarray}
    V_{\alpha}^{iso}=\frac{D}{2\rho_{0}}
    {(\rho_{\alpha}^{n}-\rho_{\alpha}^{p})^{2}}
\end{eqnarray}
The single particle potential is obtained by writing 
$U_{\alpha}^{n}\!\equiv\!\partial
V_{\alpha}^{iso}/\partial \rho_{\alpha}^{n}$ for neutrons and the same with $p$
replacing $n$ for protons.
In the above, $\rho_{\alpha}^{n}$($\rho_{\alpha}^{p}$) is the neutron(proton)
density at grid site $\alpha$.  We will also consider one more potential
which is a simplified version of the MDYI potential in which any momentum
dependence is suppressed. We refer to this potential as ``H'' with a
compressibility of K$\!=\!380$ MeV (see Table \ref{table1}). For detailed comparisons of these and other mean
field potentials as far as flow observables are concerned, the reader is 
referred to \cite{zhang}.

The initialization of the test particles in phase space is done
following usual techniques \cite{bdg88}.
Using the two parameter sets described in the previous section,
the binding energy per nucleon was calculated for nuclei in the mass
range 
$A:4\!\rightarrow\!208$. We find the difference between TPV and LHV to be
negligible for a momentum-independent potential.
For this case very good agreement
is obtained with the Weizs\"acker semi-empirical mass 
formula over the entire mass
range studied. Parameter set {\rm II} ($\delta x=0.75$ fm) deviated by no more than
about $0.5$ MeV from the mass formula. Parameter set {\rm I} consistently gave about
$1$ MeV larger binding energy per nucleon. For the momentum-dependent 
case however, both TPV
parameter sets deviated substantially from the binding energy curve for
$A<20$, giving up to $8$ MeV per nucleon too large a binding energy. Only
parameter set {\rm I} approached the binding energy curve at large $A$. Parameter set
{\rm II} gave too little a binding energy per nucleon 
($\sim6$ MeV) for $A=208$. 
For the LHV method,
the shape of the binding energy curve was reproduced for both 
parameter sets,
however parameter set~{\rm I} gave too large a binding energy per
nucleon for 
low mass nuclei
($\sim10$ MeV for $A=10$). Parameter set {\rm II} performed well, only 
giving 
$\sim\!0.5\!\rightarrow\!1$ MeV more binding than the mass formula.

Recently, it has been shown that the presence of a ``neutron skin'' should be
taken into account in low energy heavy ion collisions~\cite{b_li_2,sobotka}.
Furthermore, if we wish to reproduce as well as possible the nuclear ground
state it is desirable to be able to reproduce a neutron skin.
A recent relativistic mean field calculation by Warda~\cite{m_warda} has
given a parameterization of the neutron and proton radii in heavy nuclei ($A\!>60\!$). We used this parameterization to specify the initial 
neutron and proton
radii.
For nuclei with mass numbers smaller than $A\!=\!60$ we used the
method employed by Sobotka~\cite{sobotka} to generate neutron skins.

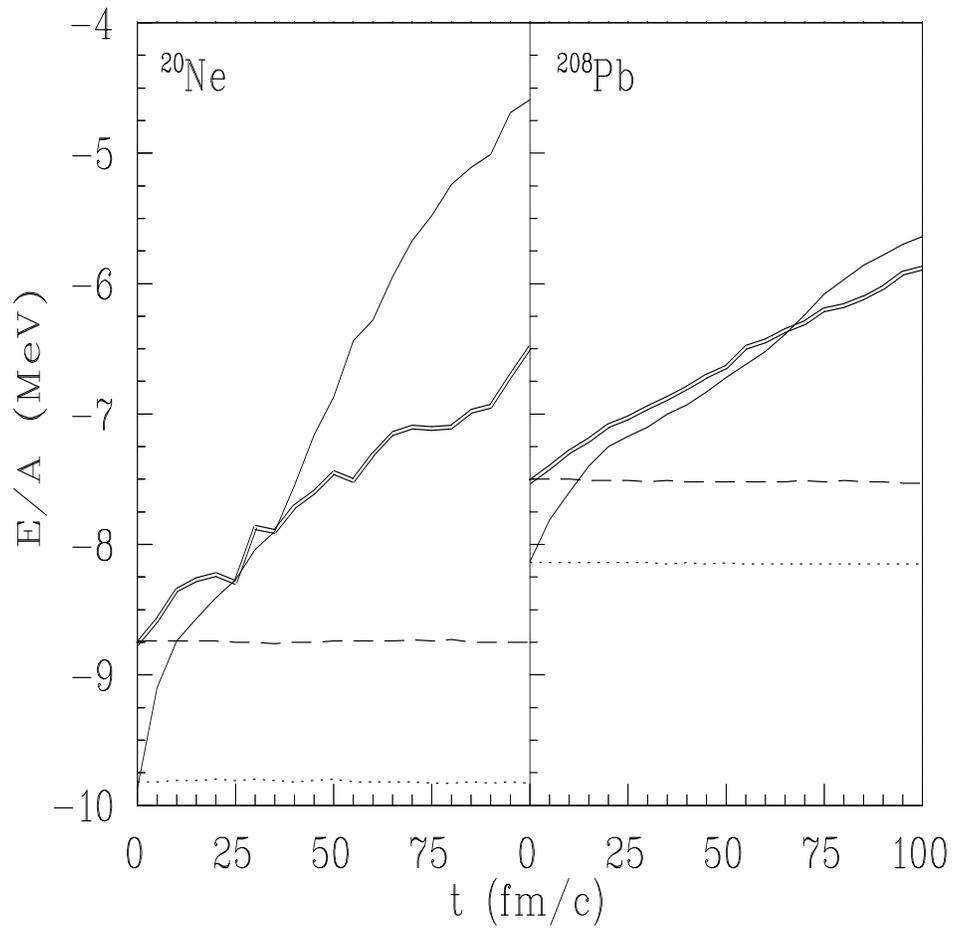
\begin{figure}[b!]  
\caption{Time evolution of the total  energy per nucleon in $^{20}$Ne and
$^{208}$Pb for a momentum-independent (K=380 MeV) nuclear mean field. 
The results for four
distinct calculations are shown. Coulomb and Isospin effects are 
included in the mean
field. The following situations were considered: TPV [n = 1, $\delta$x
= 1.5 fm] (solid line), TPV [n = 2, $\delta$x = 0.75 fm] (double
line), LHV [n = 1, $\delta$x = 1.5 fm] (dotted line), and LHV [n = 2,
$\delta$x = 0.75 fm] (dashed line). }
\label{be_t_momind}
\end{figure}

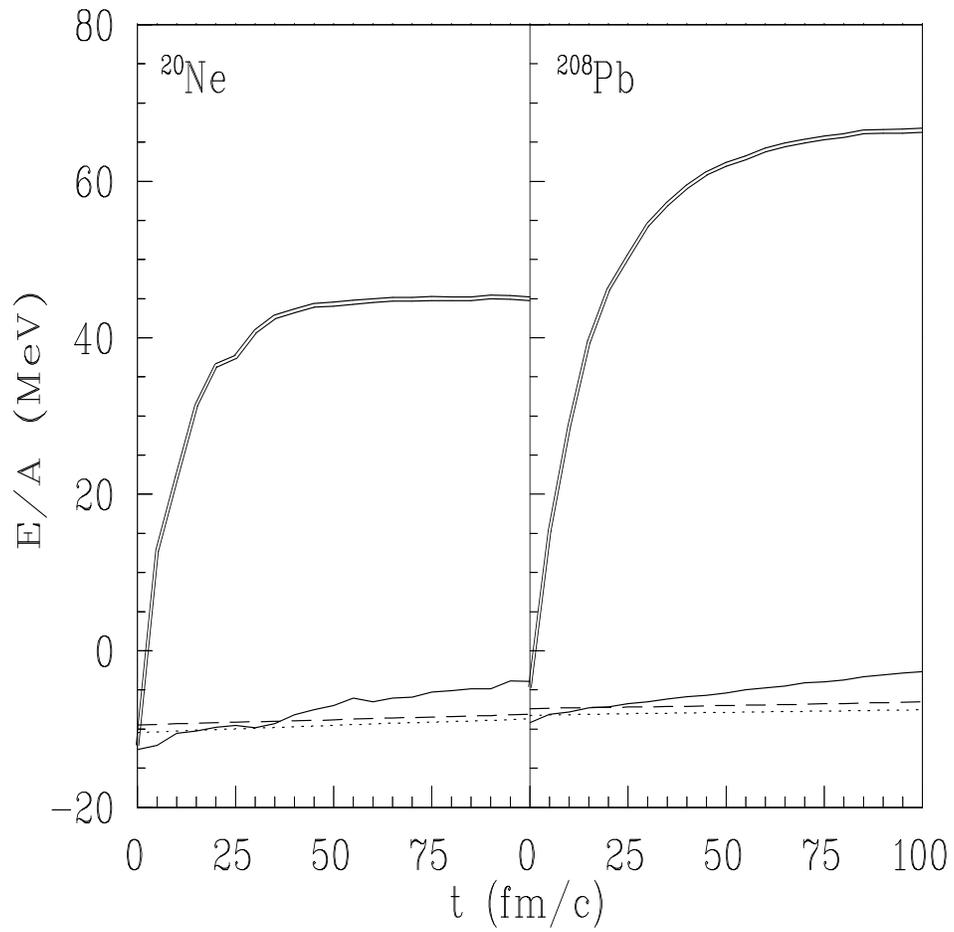
\begin{figure}[b!] 
\caption{Same as figure~\ref{be_t_momind} but with a momentum-dependent mean
field. The momentum dependence used here is of the MDYI type with a
compressibility of K=215 MeV.}
\label{be_t_momdep}
\end{figure}

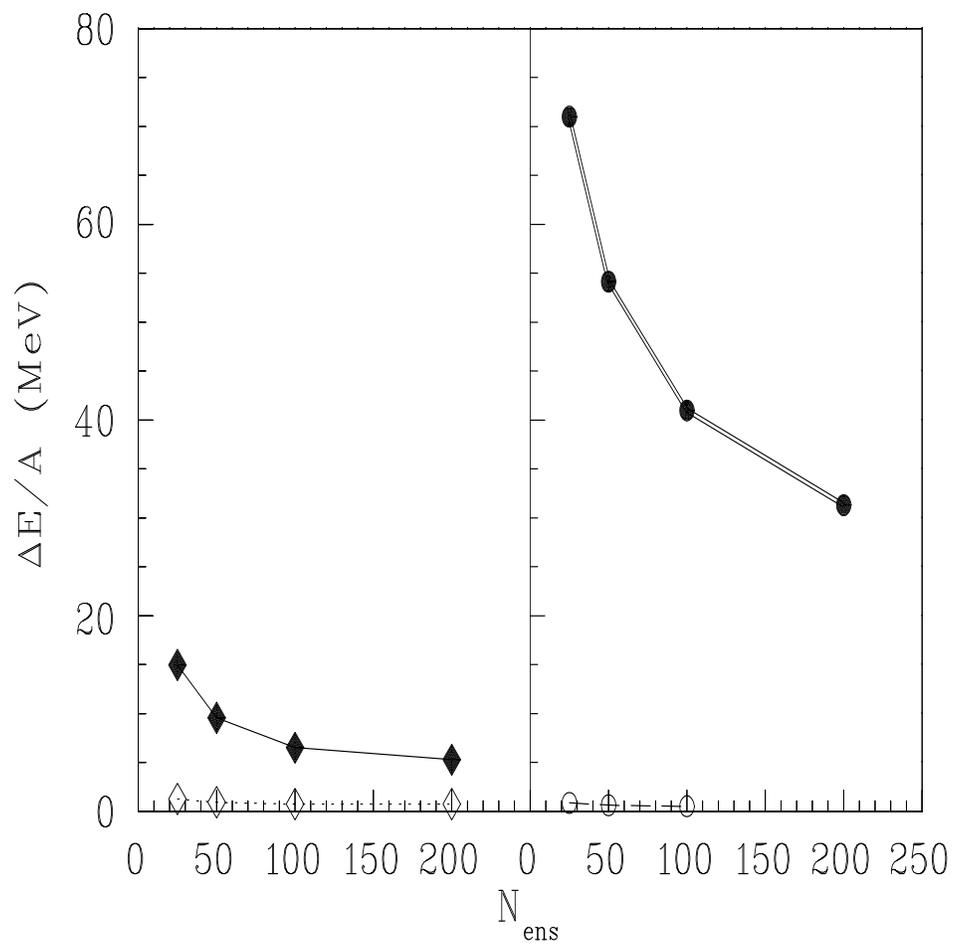
\begin{figure}[b!] 
\caption{Energy gain for a single Pb nucleus in the TPV and LHV
methods. The energy gain is plotted as a function of $N_{ens}$. Both parameter
sets considered in this work are shown and all calculations here are for the MDYI
momentum-dependent mean field potential including Coulomb and isospin effects.
The energy gain is after $100$ fm/c. Note that due to the large time required
for the momentum-dependent LHV simulation with $N_{ens}=200$, that data point
is not shown. All lines have the same meaning as those in 
figure~\ref{be_t_momind}.}
\label{e_con_n}
\end{figure}

Another test for ensuring a close approximation to ground state nuclei is the
nuclear stability as a function of time.
The total energy per nucleon for 
a small ($^{20}$Ne) and a large ($^{208}$Pb) nucleus with the two parameter
sets for the lattice spacing/form factors was investigated. The 
results are displayed in
Figs.~\ref{be_t_momind} and~\ref{be_t_momdep}. From these figures, one 
sees that
the TPV method suffers from an energy gain for both $n\!=\!1$ and
$n\!=\!2$ ($\sim2.5\ (1.6)$ MeV/nucleon at $t\!=\!100$ fm/c for 
$n\!=\!1\ (2)$) for the heavy Pb nucleus. Note that the
energy gain is less severe for $n\!=\!2$ as well as for the heavier nucleus
with a momentum-independent potential, see Fig.~\ref{be_t_momind}.
On the other hand, the LHV method shows that energy conservation is almost
complete 
(gain of $\sim10\ (30)$ KeV/A at $t\!=\!100$ fm/c for $n\!=\!1\ (2)$) for 
both parameter sets.
When we turn to the momentum-dependent case
we see the drastic difference between the two parameter sets in the TPV
method. Both suffer from energy nonconservation ($\sim6.5\ (71)$ MeV/nucleon 
at $t\!=\!100$ fm/c for $n\!=\!1\ (2)$), however, for 
$n\!=\!2$, $\delta x\!=\!0.75$ fm (double line),
the nucleus is highly unstable and quickly gains energy.
The LHV method on the other hand
shows only a very slight energy gain ($\sim730\ (860)$ KeV/A at 
$t\!=\!100$ fm/c for
$n\!=\!1\ (2)$, respectively). The numbers above are summarized in Table 
\ref{table3}.
We find that the amount of energy nonconservation is strongly
dependent upon $N_{ens}$ in the TPV method whereas the LHV method shows only
a slight dependence on $N_{ens}$. This is illustrated in
Fig.~\ref{e_con_n}. For this reason alone, it is safer to use the
LHV solution as any dependence on $N_{ens}$ as seen in the TPV case is
in some sense  spurious. Note that we
could in theory (and practice) push the limits of the TPV method by greatly
increasing $N_{ens}$.

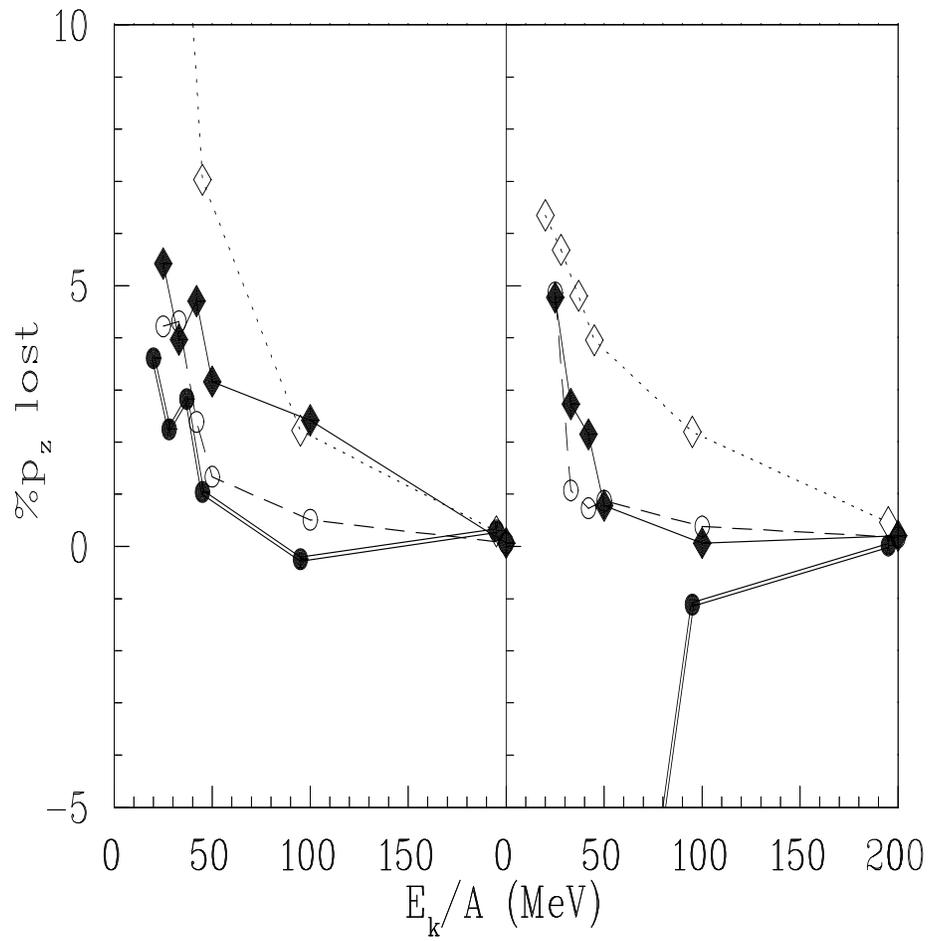
\begin{figure}[b!] 
\caption{Percentage of lost forward momenta
for a single $^{40}$Ca nucleus moving at lab energies of
$E_{k}/A:25\rightarrow200$ MeV. The results obtained with a 
momentum-independent
(K=380 MeV) and a momentum-dependent (MDYI, K=215 MeV) nuclear 
mean field are shown in the left and right panel, respectively.  Note that
for the momentum-dependent case, the TPV solution with n=2 and 
$\delta x=0.75$ fm (double line)
suffers from an energy gain large enough to overwhelm 
lattice friction effects.  All lines have the same meaning as those in 
figure~\ref{be_t_momind}.}
\label{friction_e}
\end{figure}

All calculations so far have been done on a finite configuration 
space grid. It is well known that calculations of these type 
generally break Galilean invariance and thus do not strictly 
conserve momenta~\cite{momcon_latt}. In short, a ``lattice friction''
is generated, leading to momentum nonconservation. We have 
investigated this phenomena for a single $^{40}$Ca
nucleus moving with lab energies $E_{k}/A\!:25\!\rightarrow\!200$ MeV.
At each value of the bombarding energy, the
nucleus was allowed to traverse the grid.  The momentum conservation 
results for both momentum-independent and momentum-dependent mean fields in
the TPV and LHV methods for both parameter sets are displayed in
Fig.~\ref{friction_e}. From these figures we see that when 
$n\!=\!1$, the LHV method (dotted line) is 
more susceptible to lattice friction than the TPV method (solid line) for 
all studied potentials. 
The vulnerability of the LHV to lattice friction had been
previously observed \cite{lenk_pand}. 
For $n\!=\!2$ the LHV (dashed line) and TPV (double line) give 
comparable results with a momentum-independent potential. The $n\!=\!2$ 
TPV fails
with the momentum-dependent interaction. 
Both panels of this figure show that the relative momenta
loss increases as the bombarding energy decreases. The kinks at low
energy are due to fluctuations. 
In the next section we will specify which parameter set we shall use with a
given potential, for optimal results in energy and momentum
conservation. 

\section{Nuclear Collisions in the Vlasov Limit}
The last sections elucidated the differences between the TPV and LHV approach
for the initial as well as the time evolved state of a single nucleus.
However, in heavy ion physics,
it is the {\em interaction} of two nuclei that is of interest.
So now the question arises as to how the two methods differ when we are
dealing with a system of colliding nuclei. For the moment we are 
concerned  with 
the integrity of the mean field and we need only consider the Vlasov 
limit of our model: we presently neglect hard
nucleon-nucleon scattering (see next section). We have 
investigated the time evolution
of the total centre of mass energy of a light ($^{20}$Ne+$^{20}$Ne) and a heavy
($^{208}$Pb+$^{208}$Pb) system, at bombarding energies from 25~MeV/nucleon to
400~MeV/nucleon. 
Both momentum-independent and momentum-dependent
nuclear mean fields were considered. 
We found effects similar to that which were observed when
considering only a
single nucleus. Namely, TPV suffers from energy non-conservation for both
parameter sets ($n\!=\!1$ and $n\!=\!2$), while the LHV method is much 
more stable (see below).
The collisions considered here were at zero impact parameter.

We will report on the net energy change at different
incident lab bombarding energies for the symmetric Pb system with a
momentum-dependent mean field potential. 
For all cases, the energy variations increase as the beam
energy decreases. 
For the LHV cases ($n$~=~1 and 2) no
more than $\sim\!1$ MeV/nucleon was gained in the entire 
simulations at $E_k$/A~=~50~MeV. 
This compares to
$\sim\!9$ MeV/nucleon for TPV ($n\!=\!1$) and $\sim\!80$ MeV/nucleon for 
TPV ($n\!=\!2$). In general, the
LHV method does a much better job of conserving energy. Indeed, for the better
TPV case the energy gain is already comparable to the binding 
energy per nucleon.
At kinetic energies per nucleon of around 200 MeV 
and higher, LHV yields a total
energy variation less than 0.3 MeV/nucleon.

In summary of this section and of the previous one, the TPV 
method is satisfactory
only with parameter set {\rm I} for momentum-dependent potentials. 
The LHV method
performs quite well with both momentum-independent and well as
momentum-dependent potentials in terms of energy conservation, but 
parameter set {\rm I} suffers
from lattice friction. For these reasons, we will only consider from here
on the
TPV method with parameter set {\rm I} and the LHV method with parameter set {\rm II} (see Table \ref{table1}).

\section{The inclusion of hard scattering}
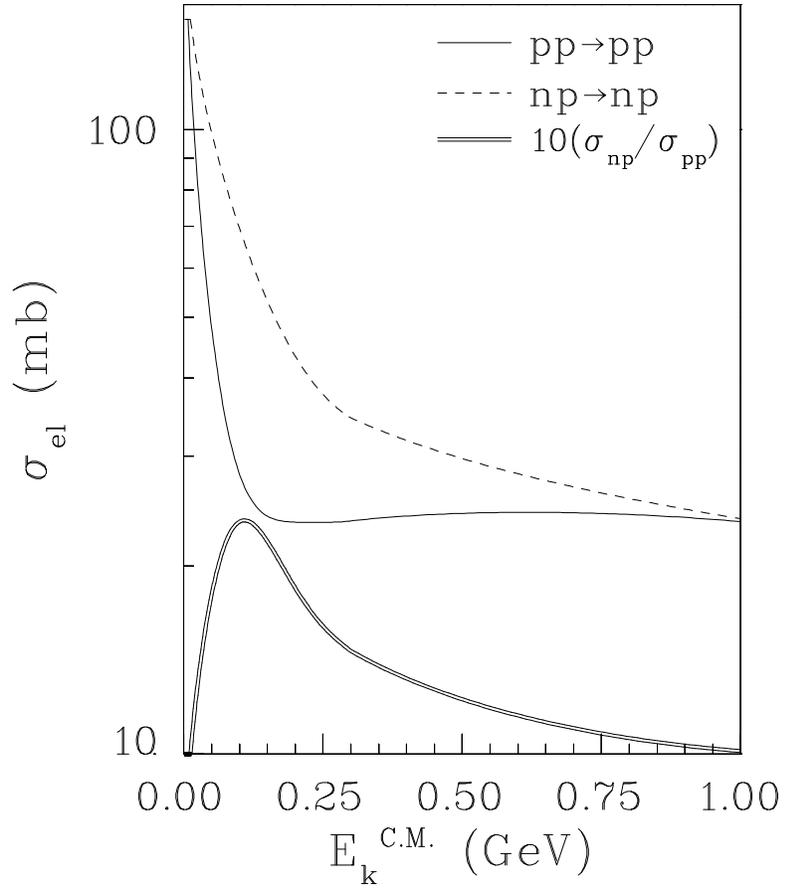
\begin{figure}[b!] 
\caption{Parameterization of the free space isospin 
dependent elastic cross section used for binary
$pp(nn)$ and $np$ collisions. Note that the isospin asymmetric 
channel ($np$) can have a
cross section of about $2.4$ time the isospin symmetric 
channel ($pp$ or $nn$). These
parameterizations are taken from Ref.~\protect\cite{j_cugnon}.}
\label{sigma_cugnon}
\end{figure}

So far we have discussed differences arising in the TPV and LHV methods
that affect the evolution of the
 mean field for both single and interacting nuclei.
Individual nucleons can also collide with each other 
({\it i.e.} undergo ``hard scattering'').
In order to include the hard scattering effects we 
require {\it a priori} the nucleon-nucleon scattering cross section. In
this work we will consider only elastic collisions. A recent 
parameterization of the
nucleon-nucleon cross section which includes isospin has been given by
Cugnon, L'H\^ote and Vandermeulen~\cite{j_cugnon}.
This cross section is an improvement over previous 
parameterizations \cite{Cugnon2} which have been used in 
BUU~\cite{zhang,bdg88} and QMD~\cite{AIC1,AIC2} calculations.
We present this cross section in Figure~\ref{sigma_cugnon}. 
From this figure ones sees that the isospin asymmetric channel can be as large
as $2.4$ times the isospin symmetric channels. We also note that for
collisions with kinetic energy equal to or smaller than  
$E_k^{\rm cm} \sim\!10$~MeV, the scattering cross
section has been set to  a constant value of $150$ mb.
We find that Pauli blocking of the final
states prohibits the majority ($>\!99\%$) of collisions with centre of mass 
energies below this value at the bombarding energies under study here. The 
energy integrated Pauli blocking efficiency for both light and heavy
ground state nuclei is found to be $\sim95\%$.

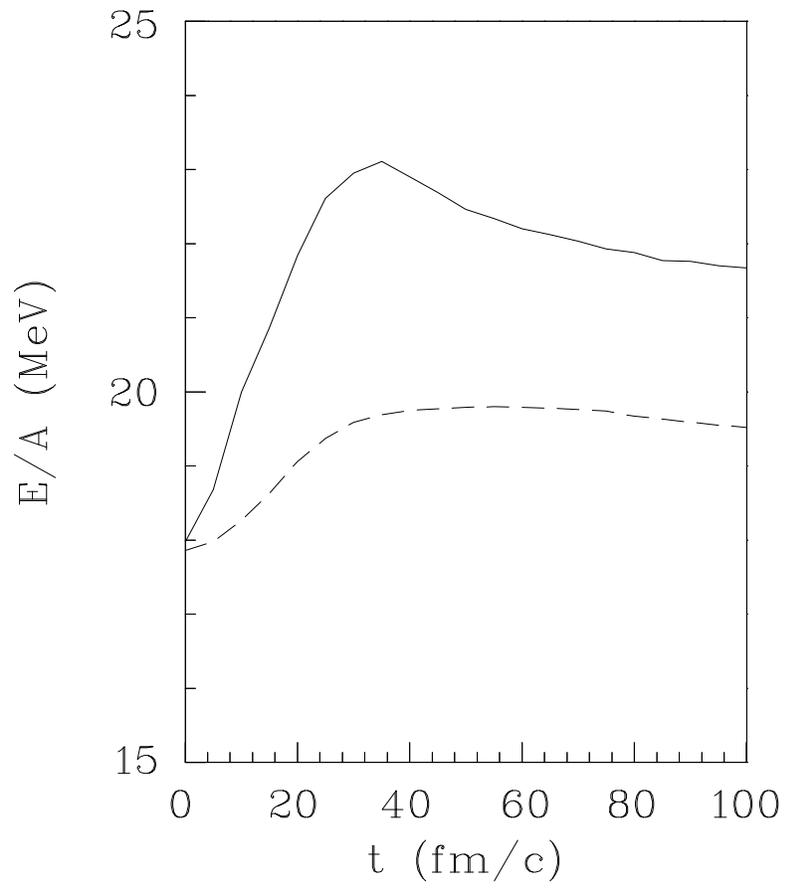
\begin{figure}[b!]
\caption{Time evolution of the total energy per nucleon in the nucleus-nucleus
centre of mass frame for $^{208}$Pb+$^{208}$Pb collisions at
a lab bombarding energy of $100$ MeV/A.
Coulomb and Isospin effects as well as hard nucleon-nucleon scattering have been
included. The results shown are for a
momentum-dependent MDYI nuclear mean field of compressibility K=215 MeV. 
The solid line is the TPB result and the dashed line is the LHB result.}
\label{e_t_momdep_boltzmann}
\end{figure}

It is well known that the inclusion of hard scattering within the framework of
a momentum-dependent mean field potential can introduce energy
nonconservation~\cite{gale_sdg}.
In addition, the inclusion of nucleon-nucleon
collisions tends to stop colliding nuclei (nucleons from both nuclei tend to
pile up around the interaction zone).  Thus, the properties of the mean field
and those of collisions add up nonlinearly. This will be
important for our comparisons of TPV and LHV since it has already been shown
that TPV does not properly handle the mean field. For these reasons, we will
reexamine energy conservation in collisions of nuclei in both the TPV and LHV
methods this time with hard nucleon-nucleon scattering present.
 Note that these ``new'' comparisons will be renamed TPB
(test-particle-Boltzmann) and LHB (lattice-Hamiltonian-Boltzmann).
Figure ~\ref{e_t_momdep_boltzmann} shows the growth of total centre of
mass energy with two colliding Pb nuclei at $E_{k}/A\!=\!100$ MeV. 
From this we see that both the TPB and LHB method now suffer from 
energy nonconservation.
However, the difference is that the energy gain in LHB is 
predominantly from collisions,
while the energy non-conservation in TPB is from both collisions and 
mean field effects.

Using MDYI we note that with the TPB method, as the
bombarding energy per nucleon decreases down to some 
critical value($\sim\!100$ MeV), the
total energy gain also decreases. Below this critical bombarding energy, the
system suffers from an energy gain which is very roughly inversely 
proportional to the lab bombarding energy.
Indeed, at low $E_{k}$, the energy gain per nucleon grows as
$E_{k}$ decreases and can be $\approx$ 6 MeV for a system of
colliding Pb nuclei at $E_k$/A~=~50 MeV. 
Since the hard nucleon-nucleon collisions and the mean field
interact highly nonlinearly it is difficult to clearly isolate the
cause of the energy gain. 
However, we do know that it should come mostly from the 
mean field propagation as most nucleon-nucleon collisions are Pauli blocked.
With LHB, on the other hand, the total energy gain gradually
decreases with decreasing $E_{k}$ and eventually levels-off to the energy 
gain seen in Vlasov
solutions ({\it i.e.} in LHV). This is just what one would again 
expect from Pauli blocking.
For the practical application we have in mind in this work, we deem the energy
conservation of the LHB solution satisfactory: for Pb + Pb at 50 MeV/nucleon,
we only gain $\approx$ 1 MeV/nucleon.  

\section{Data Comparison}
In this section we perform several comparisons of the TPB/LHB results with
recent experimental measurements on nuclear stopping. The experiment was
performed at the Michigan State University K1200 cyclotron and consisted of
bombarding beams of $^{40}$Ar on targets of Cu,Ag and Au. The beam
energy per nucleon was in the range 
$8\!\rightarrow \!115$ MeV.
A portion of this experiment involved
identifying the heavy final state remnant's mass and velocity. The ratio
$v_{\parallel}/v_{cm}$ measured in the lab frame represents the stopping
power of the above nuclear reactions. The ratio $v_{\parallel}/v_{cm}\sim\!1$
indicates large stopping and partial nuclear fusion while the ratio
$v_{\parallel}/v_{cm}\!\sim0$ indicates less stopping. Here, $v_{\parallel}$ is
the longitudinal velocity of the heavy remnant in the lab frame and
$v_{cm}$ is the velocity of the centre of mass of the projectile + target
system in the lab frame.

Before we discuss the comparisons we
will first describe some relevant details of the simulations.
We generated the nuclei with neutron and proton radii given by
the methods of Warda~\cite{m_warda} and Sobotka \cite{sobotka}. 
Note that the asymmetry parameter for Ar is small so we do not
get a significant skin for this nucleus. The gold nucleus 
on the other hand has a large
asymmetry parameter, thus the inclusion of a neutron skin is desirable. 
Next, the nuclei were boosted towards each other on Rutherford trajectories at
the appropriate beam energy.  The simulations were
run until a single large remnant was well separated 
from all other ``fragments''. 
Simulation times ranged from $t\!:325\!\rightarrow \!225$ fm/c for 
$E_{k}/A\!:~\!20\!\rightarrow \!120$ MeV, respectively.
The next step in identifying the remnant was achieved by calculating the
single particle total energy in a local rest frame. See
reference~\cite{pan_dan}, for example. A nucleon within the vicinity of the
centroid of the large remnant was considered bound to this remnant only if the
total energy of the nucleon in a local rest frame was negative. 
This allowed for a nice
identification of the heavy remnant as we shall see.
Finally, the total mass and lab frame 
velocity was then deduced from this remnant.

We considered in turn the momentum-independent and 
momentum-dependent potentials
discussed in this work. Furthermore, a second momentum-dependent
potential was considered here. We refer to this potential as GBD. Details can
be found in reference~\cite{zhang}. We find that the energy and momentum
conservation with this potential to be similar to that found in the MDYI
investigations of the previous sections.
The potential energy density for this potential reads as follows (see
also Table \ref{table1}):
\begin{eqnarray}
        V(\vec{r})=
        \frac{A}{2}\frac{\rho(\vec{r})^{2}}{\rho_{0}}
        +\frac{B}{\sigma+1}\frac{\rho(\vec{r})^{\sigma+1}}{\rho_{0}^{\sigma}}
        +\frac{C\Lambda^{2}\rho(\vec{r})}{\rho_{0}}\int d^{3}\!p
        \frac{f(\vec{r},\vec{p})}
             {\Lambda^{2}+(\vec{p}-<\!\vec{p}\! >)^{2}}.
\end{eqnarray}

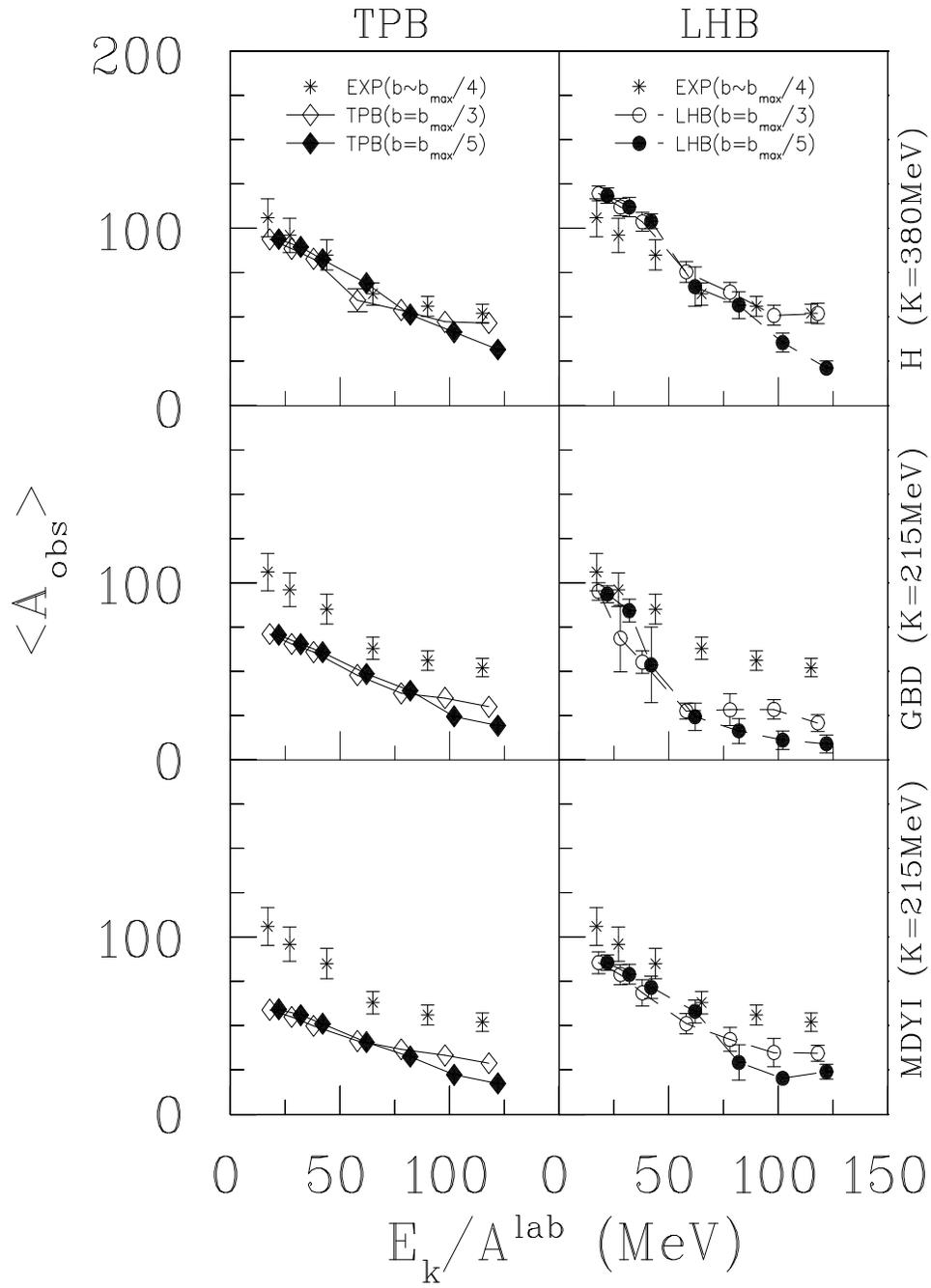
\begin{figure}[b!]
\caption{Mass number of the heavy remnant from experiment and the TPB(left
panels) and LHB(right panels) simulations for the Ar+Ag system.
The error bars on the calculated points are one standard deviation
statistical errors. The error bars on the experimental points are about 10\%
the value of the data point~\protect\cite{priv_sun}.
The top two panels represent calculations with a momentum-independent
potential. The middle two are with the GBD momentum-dependent interaction and
the bottom two are with the MDYI momentum-dependent interaction.}
\label{mass}
\end{figure}

Figure~\ref{mass} shows the results from the MSU experiment as well as the
TPB/LHB simulations for three different nuclear potentials. The
momentum-independent
potential referred to as ``H'' has a compressibility of
K$\!=\!380$ MeV. Both the GBD and MDYI potentials have compressibilities of
K$\!=\!215$ MeV. The experimental multiplicity gate corresponds to an impact
parameter $b\!\sim \!b_{max}/4$~\cite{dan_calc}. The simulations were run for
two impact parameters of $b\!=\!b_{max}/3$ and $b\!=\!b_{max}/5$ in an attempt
to bracket the data.

Let us first consider the TPB momentum-independent
calculations. We see from the figure that the remnant masses are
reproduced within experimental uncertainty. On the other hand, the LHB result
indicates that the trend is reproduced but the magnitude is slightly
overestimated at low $E_{k}$. One might be
tempted to naively assume that the larger mass seen in LHB (over that of TPB)
can be attributed to the energy gain ({\it i.e.} increased 
nucleon evaporation) in TPB which is absent in LHB. However,
when we turn to the momentum-dependent results (GBD) it is clear that this is
not the case there. The TPB results again underestimate the remnant mass and
roughly reproduce the data trend. LHB on the other hand reproduces neither the
trend nor the magnitude of the data. The GBD/LHB result shows that the
remnant mass drops rapidly for incident energies $E_{k}/A:
20\!\rightarrow\!60$ MeV and settles onto a plateau above $60$ MeV. Here, the
remnant masses are quite low ($A<25$). In fact, it is observed that the GBD/LHB
final state consisted of many small and equally sized remnants. As we shall
see, this leads to difficulties in determining the remnant velocity as it is
unclear which fragment represents the ``large'' remnant.  Finally, the
MDYI/TPB result is similar to GBD/TPB ({\it i.e.} magnitude 
underestimated).  On the other 
hand, the MDYI/LHB result shows
much better agreement with the data trend and only slightly underestimates the
remnant mass. From the figure it is clear that the momentum-independent TPB
gives the best overall agreement with the data. For the momentum-dependent
result, the best agreement is obtained with MDYI/LHB.

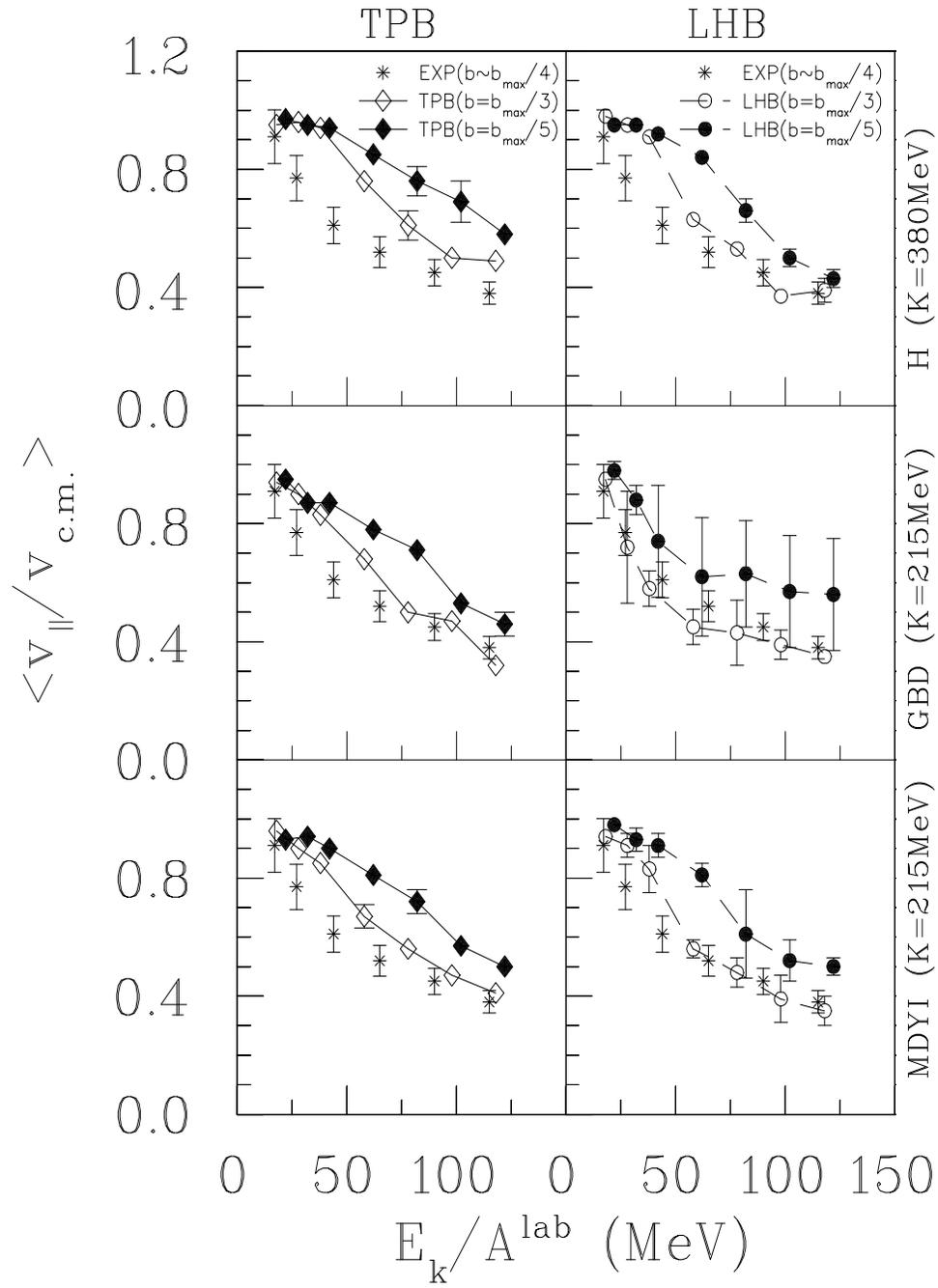
\begin{figure}[b!] 
\caption{Experimental and calculated average velocity ratios of the heavy
remnant from figure~\ref{mass}.
The error bars on the calculated points are one standard deviation
statistical errors. The error bars on the experimental points are about 10\% the
value of the data point~\protect\cite{priv_sun}.
All panels are as described in figure~\ref{mass}.}
\label{v}
\end{figure}

Next we turn to the remnant velocity distributions. Figure~\ref{v} shows the
experimental and calculated results for the three potentials and two methods
discussed previously. We see from this figure that the
momentum-independent calculations fail to reproduce the data for both 
TPB and LHB. In fact, 
for all three potentials the TPB results do not reproduce the data.
However, we have better agreement with a momentum-dependent potential.
When we turn to the LHB momentum-dependent results, it is observed that the GBD
potential just brackets the data. In fact for $b=b_{max}/3$, the data is
reproduced quite well. The MDYI also does a better job than the
momentum-independent
potential, especially at the larger impact parameter. However the 
agreement is not as good as with the
GBD potential. Note the large error bars seen in the GBD/LHB calculation.
This follows from the ambiguity encountered in determining the large 
final state remnant discussed in the previous paragraph, this 
difficulty is absent in the calculations done using other interactions.

\section{Conclusion}
In summary, we see agreement with the measured remnant mass obtained
with the LHB method, except with the GBD potential. This potential
gives poor agreement with this data. Both the 
momentum-independent and MDYI
potentials give good agreement with the data. For the velocity distributions,
the TPB method failed for all three potentials. Reasonable agreement could be
achieved only with a momentum-dependent potential in the LHB method.
Overall it appears that the MDYI interaction, used with the Lattice
Hamiltonian algorithm achieves a satisfactory general description of the
experimental data. Coupled with the reasonable ground state solutions of this
combination and with the fact that the MDYI receives support from
higher energy flow data \cite{zhang}, this situation is satisfying. 
Further note that a hard momentum-dependent interaction does not seem
to be supported by the higher energy data, nor by the data discussed in
this work \cite{mrst99}. 

We are now in
a position to consider other observables such as the balance energy
and also to try and extract information on possible in-medium variations
of the nucleon-nucleon cross sections. Those issues  are under
investigation. By now it is clear that progress in the theoretical
study of heavy ion collisions can only be achieved through the
simultaneous investigation of a collection of related physical
observables.

\acknowledgments
We are happy to acknowledge useful discussions with P. Danielewicz, 
S. Das Gupta, and R. Lacey. This work is supported in part by 
the National Science and
Engineering Research Council of Canada and in part by the Fonds FCAR of the
Qu\'{e}bec government.  One of us (D.P.) is happy to acknowledge the financial
support of McGill University through the Alexander McFee Memorial Fellowship.

\begin{table}
\caption{Parameters for nuclear matter potentials considered in this
work. All entries have units of MeV except $\sigma$ and $m^*/m$, which
are pure numbers.}\label{table1}
   \begin{tabular}{ccccccccccc}
potential	&A&B&$\sigma$&C&$\Lambda$&$m^{*}/m$&$U(\rho_{0},p_{f})$
						   &$U(\rho_{0},0)$
                                                   &$U(\rho_{0},\infty)$
						   &K\\ \hline
H	&-124.0&70.5&2.0&-&-&1&-53.5&-53.5&-53.5&380\\
GBD	&-144.0&203.3&7/6&-75.0&400.0&0.7&-53.3&-76.3&-1.34&215\\
MDYI	&-110.4&140.9&1.24&-64.95&415.7&0.67&-52.9&-75&30.5&215\\
	\end{tabular}
\end{table}

\begin{table}
\caption{Parameter sets used for the single-particle equations of
motion.}\label{table2}
   \begin{tabular}{cccc}
parameter set	&$n$		&$\delta x [ fm]$	&$N_{ens}$\\ \hline
{\rm I}		&1		&1.50			&100\\
{\rm II}		&2		&0.75			&25\\
   \end{tabular}
\end{table}

\begin{table}
\caption{Energy gain for a single Pb nucleus after $t=100$
fm/c. All entries have units of MeV.}\label{table3}
   \begin{tabular}{ccccc}
parameter       &\multicolumn{4}{c}{$\Delta E [ MeV]$}\\ \cline{2-5}
set      	&\multicolumn{2}{c}{TPV}&\multicolumn{2}{c}{LHV}\\ \cline{2-5}
           &mom.ind.   &mom.dep.   &mom.ind.   &mom.dep\\
   \hline\hline
{\rm I}               &2.5       &6.5        &0.01       &0.73\\
   \hline
{\rm II}               &1.6       &71     &0.03       &0.86\\
   \end{tabular}
\end{table}

\end{document}